\documentclass[10pt, conference]{IEEEtran}
\IEEEoverridecommandlockouts
\usepackage{blindtext, graphicx}
\usepackage{listings}
\usepackage[linesnumbered,ruled,english]{algorithm2e}
\usepackage{graphicx}
\usepackage[english]{babel}
\usepackage{cite}

\usepackage[utf8]{inputenc}
\usepackage{subfigure}

\usepackage{fixltx2e}
\usepackage{color}
\usepackage{algpseudocode}
\usepackage{verbatim}
\usepackage{balance}

\title{FedSA: Accelerating Intrusion Detection in Collaborative Environments with\\ Federated Simulated Annealing}

\author{
\IEEEauthorblockN{Helio N. Cunha Neto\IEEEauthorrefmark{1}, Ivana Dusparic\IEEEauthorrefmark{2}, Diogo M. F. Mattos\IEEEauthorrefmark{1}, and 
Natalia C. Fernandes\IEEEauthorrefmark{1}}

\IEEEauthorblockA{\IEEEauthorrefmark{1}MídiaCom - PPGEET/UFF\\
Universidade Federal Fluminense - Niterói, Brazil}
\IEEEauthorrefmark{2}School of Computer Science \\
Trinity College Dublin - Dublin, Ireland
}

\begin{document}

\maketitle
\thispagestyle{empty}
\pagestyle{empty}

\begin{abstract}
Fast identification of new network attack patterns is crucial for improving network security. Nevertheless, identifying an ongoing attack in a heterogeneous network is a non-trivial task. 
Federated learning emerges as a solution to collaborative training for an Intrusion Detection System (IDS). The federated learning-based IDS trains a global model using local machine learning models provided by federated participants without sharing local data. However, optimization challenges are intrinsic to federated learning. This paper proposes the Federated Simulated Annealing (FedSA) metaheuristic to select the hyperparameters and a subset of participants for each aggregation round in federated learning. FedSA optimizes hyperparameters linked to the global model convergence.
The proposal reduces aggregation rounds and speeds up convergence. Thus, FedSA accelerates learning extraction from local models, requiring fewer IDS updates. The proposal assessment shows that the FedSA global model converges in less than ten communication rounds. The proposal requires up to 50\% fewer aggregation rounds to achieve approximately 97\% accuracy in attack detection than the conventional aggregation approach.

\end{abstract}

\section{Introduction}

The number of connected devices on the Internet will reach 27 billion by 2027\footnote{Available at https://www.ericsson.com/en/reports-and-papers/mobility-report/dataforecasts/iot-connections-outlook. Accessed in 10/12/2021}. Different device types connect to various access networks, leading to the Internet of Everything, a heterogeneous and mutable networking environment~\cite{Mattos:2019}. Due to the Internet growth and heterogeneity, the complexity of mapping the vulnerabilities from connected devices also grows~\cite{andreoni2019}. Moreover, the spread of the attacking tools (\textit{e.g., slowloris, hping3, macof}) enables quick implementation of a network attack. 

Intrusion Detection Systems (IDS) effectively identify already-known network attacks. However, traditional IDS either deploy a signature database to recognize known attacks or analyze network behavior, searching for outliers or anomalous behaviors, which reduces the efficacy of an IDS~\cite{ALDWEESH2020}. Consequently, traditional IDS are ineffective for detecting new attack patterns. In contrast, recent Machine Learning-based IDS (ML-IDS) show a significant advance in detecting new attack patterns~\cite{andreoni2019,Sanz:2018,liu2019machine,Souza2020,Silva:2020}. Besides learning attacks on an extensive dataset, ML-IDS learns from new traffic coming from the network as long as traffic is appropriately labeled. The more data the machine learning model consumes, the more accurate it becomes~\cite{Yang2019}. Therefore, an ideal strategy is to create a model that could learn from the traffic of different networks, but it is challenging due to strict data protection laws, \textit{e.g.}, General Data Protection Regulation (GDPR).

Federated learning emerges as a feasible solution to the privacy problem in collaborative training~\cite{BrendanMcMahan2017}. Federated learning allows several participants to train a global machine learning model in a distributed fashion without data sharing. The collaborative learning approach enriches the training, building a global model based on extensive participant data. The centralized approach is less efficient as it limits the training to an offline dataset or data located on a single monitored network. In federated learning, participants train local models with their local data and share   a central server only those model parameters that create an aggregate global model based on those parameters. The aggregation server selects random participants for training at each iteration~\cite{BrendanMcMahan2017}, called aggregation round. The main goal is sharing knowledge without exposure of participants' data. 

A key challenge in designing federated learning-based IDS is that the network data used for training the global model is highly Non-Independent and Non-Identically Distributed (Non-IID)~\cite{Lim2020}. In contrast to distributed machine learning, federated learning assumes that the server does not access the participants' data. The absence of data access makes the training more difficult. The data dependence and the heterogeneous data distribution pose challenges for the optimal convergence of the trained model. The optimization issue delays the learning of new attack patterns. Furthermore, not all participants contribute to the training of the global model. Therefore, efficiently selecting participants that contribute the most to the training speeds up the global model convergence. However, selecting participants is non-trivial since data always remains private for each participant. Previous works propose the use of federated learning for intrusion detection~\cite{preuveneers2018, Nguyen2019, rahman2020internet}, but these works disregard the federated learning optimization issues, and the traditional federated learning algorithm imply delays in learning new attack patterns.


This paper proposes the Federated Simulated Annealing (FedSA), a metaheuristic for optimizing participants and hyperparameters selection in a federated learning environment. Our proposal optimizes the model hyperparameters leading to fast convergence. The optimized hyperparameters are the learning rate, local updates, and the participant selection.
To the best of our knowledge, our paper is the first to propose a federated version of the simulated annealing metaheuristic applied to the learning optimization of Federated Learning. Traditionally, discovering the best hyperparameters requires a time-consuming fine-tuning process.
The hyper-parameters selection task usually deploys a brute-force searching algorithm (\textit{e.g.} grid search, random search) to improve the model performance. Saving the time spent on fine-tuning is very important for applications like Federated IDS, where fast learning is critical.
On the other hand, our proposal provides a meta-heuristic that enhances Federated Learning performance in an autonomous and convergent way, without relying on brute-force searching algorithms. Although we focus on providing a Federated Learning approach for the IDS scenario, our proposal is the metaheuristic, which naturally applies to other scenarios.

Our proposal also halves the number of required aggregation rounds. Reducing the number of aggregation rounds decreases the use of valuable resources, such as communication and computation costs on both the server and participant sides.

We evaluate our proposal and demonstrate that the proposed FedSA outperforms the conventional federated-learning aggregation approach. We employ the CICIDS2017 as a benchmark dataset to simulate a network environment. The CICIDS2017 is a network dataset comprising normal and attack network flows samples. The results show that FedSA requires $ \approx 50\% $ fewer aggregation rounds than the conventional approach to achieve $ \approx 97\% $ accuracy. Preuveneers \textit{et al.} apply the CICIDS2017 dataset in a federated learning environment with no optimization concerns and achieved 97\% accuracy after 50 aggregation rounds~\cite{preuveneers2018}. Our proposal required five aggregation rounds to achieve the same result. In our implementation, we used FedAvg as the baseline, using multi-layer perceptron, which achieved 97\% accuracy in eight aggregation rounds.
We also evaluate the FedSA varying its hyperparameters, temperature and cooling ratio, and the result shows that the proposal is resistant to changes in bootstrap values of temperature and cooling ratio.

The remainder of the paper is organized as follows. Section~\ref{sec:related} discusses related work. We describe the federated learning in Section~\ref{sec:federated}. 
Section~\ref{subsec:fedsa} describes our proposed metaheuristic FedSA. We present simulation results in Section~\ref{sec:results}. Finally, Section~\ref{sec:conclusion} concludes the paper.

\section{Related Works} 
\label{sec:related}

Previous related research proposes collaborative learning through an IDS federation using federated learning, extending the learning perimeter~\cite{Nguyen2019, preuveneers2018, Chen2020ids, cmfl2019}. However, none of these works accelerate the convergence of the global model to speed up the learning of new attack patterns.

Nguyen \textit{et al.} propose a federated learning model for cyber-attack detection in an IoT access network. The proposed model considers a random participants' selection for each round~\cite{Nguyen2019}. IoT gateways operate as participants in federated learning, and an IoT security service provider acts as an aggregation server for models collaboratively trained. The authors evaluate the proposal in a real smart-home environment and successfully detect 95.6\% of attacks in approximately 257~ms without triggering false alarms.

Preuveneers \textit{et al.} propose to deploy blockchain technology for sharing participants' parameters in a federated IDS environment~\cite{preuveneers2018}. The proposal stores all incremental updates to the machine learning model in the blockchain ledger. Despite the advantages of shared storage, the authors identify that blockchain generates latency in training. Moreover, blockchain also introduces communication and processing overhead~\cite{Oliveira-Icin:2019}. The proposed architecture depends on the network nodes to send all their flows via an agent to the IDS analysis. The proposal requires up to 50 aggregation rounds to achieve 97\% of accuracy. Preuveneers \textit{et al.} evaluate their proposal on the CICIDS2017 dataset.

Chen \textit{et al.} propose the Federated Learning-based Attention Gated Recurrent Unit (FedAGRU)~\cite{Chen2020ids} for IDS. The authors deploy a Gated Recurrent Unit (GRU) Neural Network but replace the output layer with a Support Vector Machine (SVM). The Global aggregation is asynchronous, \textit{i.e.}, the server does not wait for all selected participants to send their parameters. The participants compare their model updates with the current global model at each global iteration by correlating the parameters. The participants only send their parameters if their update is relevant to the training. 
Mothukuri {\it et al.} also use GRU as the deep learning model for federated learning-based IDS~\cite{Mothukuri2021}. The authors proposed a federated learning approach using an ensemble to enable anomaly detection in IoT networks~\cite{Mothukuri2021}. The authors train the model against a Modbus network dataset. The proposal evaluation considers GRU and Long Short-Term Memory (LSTM), and GRU models outperform LSTM, achieving a higher accuracy rate and being computationally inexpensive. 



Rey \textit{et al.} propose a federated learning-based IDS using Multilayer Perceptron (MLP) and autoencoder models~\cite{rey2021}. The authors used the N-BaIoT, a dataset modeling network traffic of several real IoT devices while affected by malware. The authors compare centralized, distributed, and federated learning architectures. 


Unlike the previous works, our proposal is concerned with the fast convergence of the global model. The faster the convergence, the faster the model learns with the participants' data. Other works aim to solve optimization challenges from federated learning~\cite{Nguyen2021safl, smith2017}. Nguyen {\it et al.} uses Simulated Annealing to optimize federated learning training~\cite{Nguyen2021safl}. The technique is called simulated annealing-based federated learning (SAFL). The SAFL performs the local update based on the aggregation server feedback. The selected participants disturb some local parameters based on the Simulated Annealing probability at each aggregation round. The SAML avoids local optima or overfitting the local model. The authors used the MNIST, Fashion-MNIST, CIFAR10, and Google speech commands datasets for evaluations.
Smith \textit{et al.} propose MOCHA, an optimization structure for the federated environment, which allows the customization of federated learning through the learning of separated and related models for each device~\cite{smith2017}. MOCHA calibrates on the resource restrictions of a participating device, such as network conditions and CPU states of the devices. The method has verifiable theoretical convergence guarantees, but it is limited in scale to massive networks and restricted to convex objectives~\cite{smith2017}. For Evaluation, the authors used Google Glass (GLEAM), Human Activity Recognition, and vehicle sensor datasets. 

Wang et al. propose an algorithm to determine the trade-off between local updates and aggregation rounds~\cite{Wang2019}. The authors analyze the convergence boundary of federated learning based on the gradient descent from a theoretical perspective and propose a new convergence boundary. The convergence boundary incorporates the data distribution, usually unbalanced, among participants.
It is possible to determine the ideal frequency of aggregations rounds, saving computational resources. For evaluation, the authors used the MNIST dataset.



Previous works aim to use federated learning for federated IDS. However, these works do not address a crucial issue implied in the federated learning scenario, the global model convergence. Other works aim to solve the convergence issue. Nevertheless, these works focus on other application of Federate Learning, such as image processing~\cite{Wang2019, smith2017, Nguyen2021safl} or Natural Language Processing (NLP)~\cite{Nguyen2021safl}. Our proposal manages the adaptive selection of hyperparameters related to the global models' convergence and the participant selection. None of those above-mentioned works focus on providing global-model adaptive changes into hyperparameter based on previous aggregation rounds loss function values.

\section{Federated Learning}
\label{sec:federated}


The federated learning system has two main entities: the participants, who own the data, and the aggregation server, responsible for the global model creation. Let $ N = \{1,. . . , n \} $ be the set of participants, where each participant has its own private dataset $D_n, {n \in N} $. Each participant $ n $ uses his dataset $ D_{n} $ to train a local model $ w_{n}^t $, where $ t $ denotes the current aggregation round. In each aggregation round, a subset of the devices $S^{t}$,$S^{t} \in N$, is randomly selected to provide their local model's parameters to the server. Then, the aggregation server assembles all parameters from the selected participants to generate an updated global model $w_{G}^{t}$. Local models update by $ \tau $ local updates before sending the local model's parameters to the server for the global aggregation. After the global aggregation, the aggregation server sends the global model $w_{G}^{t}$ to all federation participants. An underlying assumption is that participants are well-behaved, implying they use their actual private data to train and send the proper parameters to the server.

\begin{figure}[h!]
\vspace{-3mm}
    \centerline{\includegraphics[width=.5\textwidth]{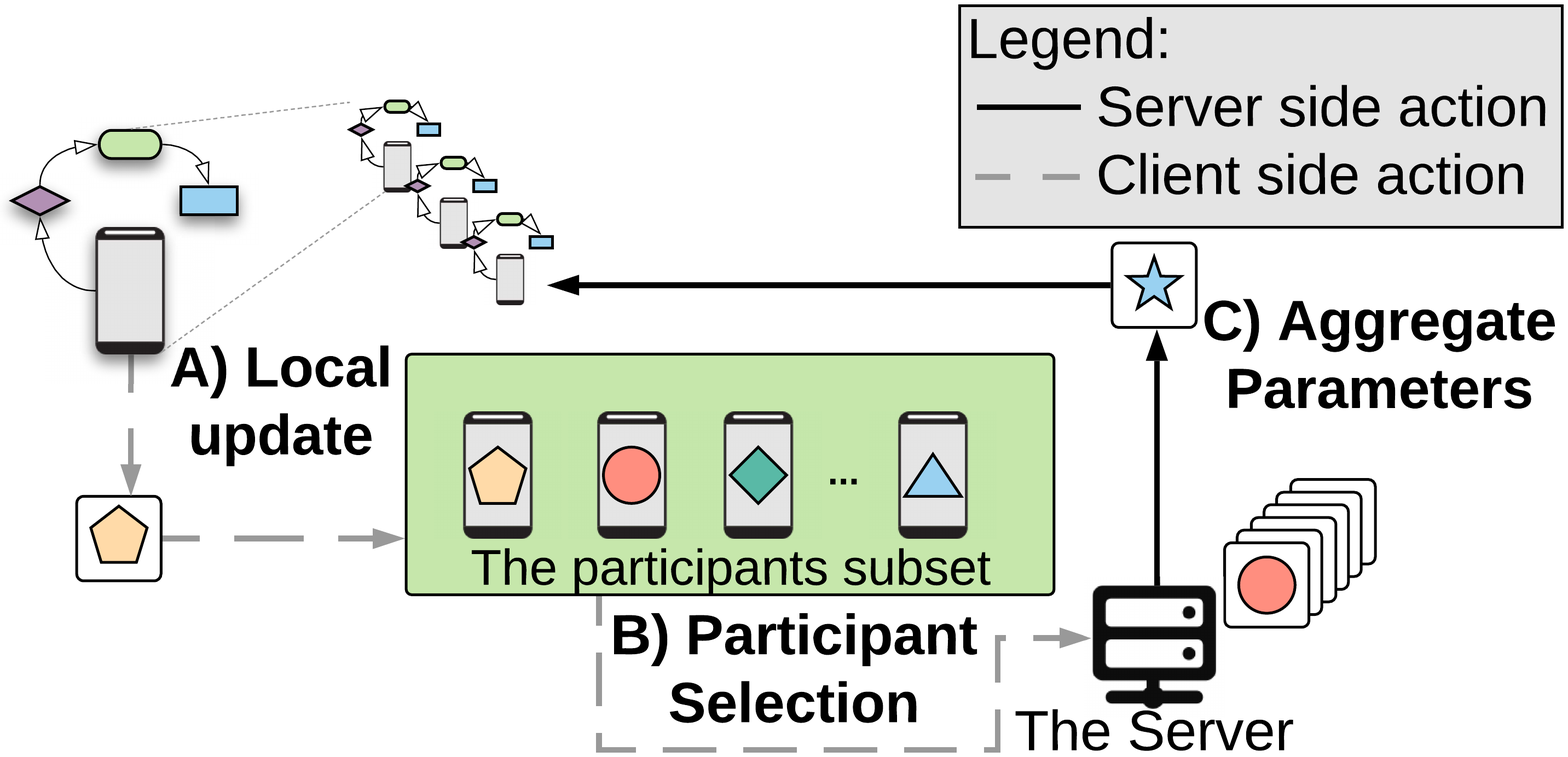}}
    \caption{The federated learning basic steps. The primary forms represent the models' parameters.} \vspace{-3mm}
    \label{fig:flsteps}
\end{figure}

The Federated Learning performs three basic steps at each aggregation rounds. The first step is called local update, the second is participant selection, and the last is global aggregation. However, an early step called Initialization occurs before the first aggregation round. In the Initialization, the server first creates the initialized global model template $ w^{0}_{G} $. This initial model is set with random parameters values, usually from a normal distribution. The server also specifies the training hyperparameters, such as learning rate ($ \eta $), local updates number ($ \tau $), and mini-batch size ($ B $). Figure~\ref{fig:flsteps} shows the federated learning three basic steps. The basic steps are described below.

\renewcommand{\labelenumi}{\alph{enumi})}
\begin{enumerate}
    
    \item \textbf{Local Update:} Based on the global model $ w^{t}_{G} $ received from the server, the selected participants use their local data and devices to update their local model parameters $ w^{t}_{n} $. The objective of the $ n^{th} $ participant in the communication round  $ t $ is to find the optimized parameters $ w^{t}_{n} $ that minimizes the local loss function $ L(w^{t}_{n}) $.
    
    \item \textbf{Participant Selection:} The server randomly selects the subset $S^{t}$,$S^{t} \in N$ of participants for training. Only the selected participants will send the local models' parameter to the server for aggregation.
    
    \item \textbf{Global Aggregation:} The server aggregates the selected participants' local model parameters and returns the updated global model $ w^{t + 1}_{G} $ back to all the participants. The server's goal is to minimize the global loss function $ L(w^{t}_{G}) $. A global aggregation occurs each aggregation round.
\end{enumerate}

The three steps repeat until the global loss function's convergence or the algorithm meets a stop condition. Federated Learning works with models based on Stochastic Gradient Descent (SGD) methods~\cite{BrendanMcMahan2017}, such as neural networks, linear regression, and support vector machines. 

Each participant dataset $D_{n} $ will have a local loss function $F_{n}(w) $.
It is important to highlight that to facilitate the visualization of the equations, $ D_{n} = |D_{n}| $, where $|$$^. $$| $ denotes the cardinality of a set and $ D = \sum _{n = 1}^{N} D_{n}$. Assuming that $ D_{n} \cap D_{n'} = \emptyset$, $\forall$ $ n \not= n' $, the global loss function for all datasets is
\begin{equation}
F(w) = \sum _{n = 1}^{N} \frac{D_{n}}{D} F_{n}(w) .    
\end{equation}

Note that $F(w)$ cannot be calculated directly without sharing information among participants~\cite{Wang2019}. The pieces of information shared by the participants are the dataset size and the local loss.
The optimization problem is then to minimize $ F(w) $, \textit{i.e.}, to find $ w^* = \arg \min F(w) $.

The first aggregation algorithm for federated learning was Federated Averaging (FedAvg)~\cite{BrendanMcMahan2017}. The FedAvg does not optimize the federated learning hyperparameters. The aggregation in FedAvg consists of a weighted sum of all parameters received from the participants. The FedAVG measures the participant relevance by the participant dataset size. The larger the participant dataset size, the greater the relevance in the aggregation. The parameters' aggregation is given by
\begin{equation}
     w^{t}_G \leftarrow\sum _{n=1}^{N} \frac{D_{n}}{D} w_{n}^{ t},
     \label{eq:fedavg}
\end{equation}
 \noindent $ w_{n}^t $ is the aggregated weight vector in the centralized server.

\section{The Proposed FedSA Metaheuristic}
\label{subsec:fedsa}

The FedSA is a Simulated Annealing (SA) variant for optimizing the federated learning hyperparameters. We have selected Simulated Annealing as the basis for our metaheuristic due to it being proven to converge to the optimum global value~\cite{convproofsa}. 
%
%
%
%
The SA metaheuristic aims to find the global minimum of a function based on the principles of statistical mechanics. SA seeks an optimal $\varsigma$ solution derived from an initial randomly generated solution. At each iteration, the metaheuristic generates a random solution $ \varsigma' $ in the neighborhood of the currently best solution known. A loss function $f(\varsigma')$ evaluates the performance of the candidate solution $\varsigma'$. In the federated learning problem, this is the global loss function. Solutions that minimize the loss function are always accepted. Otherwise, the solution $\varsigma'$ is probabilistically accepted. The acceptance probability depends on the current simulated temperature parameter $ T $ and on the degradation of $ \Delta E $ of the loss function, where  $ \Delta E $ is the difference between the value of the loss function for the current solution $ f(\varsigma) $ and for the candidate solution $ f(\varsigma') $, hence
\begin{equation}
  \Delta E = f (\varsigma') - f (\varsigma).  
\end{equation}

Therefore, the acceptance probability function is given by
\begin{equation}
 P(\Delta E, T) = \exp \left ( - \frac{\Delta E}{T} \right ),   
\end{equation}
\noindent where low temperature reduces the acceptance probability of a worse solution being the best solution. The negative value of $ \Delta E $ and $ T $ ratio in the exponential function enforces the probability decrease.
$ P (\Delta E, T) $ assesses the probability of transitioning from the current best solution $ \varsigma $ to a worse candidate solution $ \varsigma' $. The acceptance probability follows the
Boltzmann distribution~\cite{van1987simulated}. The parameter $ \alpha $, where $ 0 < \alpha <1 $, is a constant, that stands for the gradual reduction of the simulated temperature.


\begin{algorithm}[ht!] 
 	\KwIn{$ \eta $, $\tau$, $\mu$, $T_{init}$, epochs, $\alpha$ }
\KwOut{$best\_solution$}

$T\gets T_{init}$

$best\_solution \gets $ \textbf{{INIT}{$ ( \eta ,  \tau ,  \mu ) $}}

$best\_loss \gets $\textbf{{AGGREGATE}{$(best\_solution)$}}

\For{$i < epochs$}
{
    $current\_solution \gets $ \textbf{GEN\_NEIGHBOR\_SOLUTION}{$(best\_solution)$}
    
    $current\_loss \gets $\textbf{\textbf{AGGREGATE}{$(current\_solution)$}}
	
	$\Delta E\gets$ \textbf{{$current\_loss$}} $-$ \textbf{{$best\_loss$}}
	
	$p = \exp(-\frac{\Delta E}{T})$
	
	\If{$best\_loss < current\_loss$}{
	$best\_solution \gets current\_solution $
	
	$best\_loss \gets current\_loss$
	}
	
	\ElseIf{\textbf{unif}{$ (0,1) $} $ < $ 
	{$ p $}
	}{
	$ best\_solution \gets current\_solution $
	
	$best\_loss \gets current\_loss$
	
	$ T\gets $ \textbf{\textbf{COOL}{$ (T,\alpha) $}}
	}
	
	$best\_loss\_test \gets $ \textbf{\textbf{AGGREGATE}{$(best\_solution)$}}
	
	\If {$best\_loss\_test < best\_loss$}{
	$best\_solution \gets $ \textbf{\textbf{INIT}{$ ( \eta ,  \tau ,  \mu ) $}}
	}

}
\textbf{return} $ best\_solution $
 
 	\caption{Federated Simulated Annealing Metaheuristic.} 
 	\label{algo:FedSA} 
\end{algorithm}

The proposed FedSA metaheuristic aims to find the best combination of participant selection, learning rate, and local updates for each aggregation round, as shown in Algorithm~\ref{algo:FedSA}. Therefore, the FedSA expects as input the set of participants IDs ($ \mu $), the learning rate values limits ($ \eta $), and the number of local updates values limits ($\tau $). We model a solution $ \varsigma $ as a tuple containing local updates, the learning rate, and the selected participants' indexes (IDs).

The $ \eta $ tuple comprises the interval between the smallest and the highest value that the learning rate may assume during training and the $ \tau $ tuple comprises the interval of possible values for the number of local updates. The $ \mu $ contains a list with the IDs of all participants. 
Every participant associates with an index, a unique identifier, and indicates the sequence that participants joined the federation. 
FedSA bootstrapping is to perform the hyperparameter selection at random (line 2), \textit{i.e.}, the participants' subset selection, the learning rate, and local updates are random. The subsequent selections are neighboring solutions to the best solution generated in each aggregation round. Other parameters of the algorithm are the initial temperature ($ T_{init} $), the maximum number of epochs ($ epochs $), and a cooling value ($ \alpha $). A first aggregation round evaluates the initial random solution (line 3), which returns the global loss. Every solution is evaluated according to its global loss. After the aggregation, the algorithm starts the loop (line 4) to find neighbor solutions of the current best solution.

The federated-learning hyperparameters optimization scenario differs from the traditional NP-hard problems solved with SA. In traditional NP-hard problems, \textit{e.g.}, the traveling salesman, the loss of a given solution $ \varsigma $ is the same regardless of the algorithm epoch. However, in the federated learning scenario, the solution $ \varsigma $ loss in iteration $ i $ may differ from the same solution loss in iteration $ i + 1 $. The training process is stochastic since the same solution returns distinct losses. Hence, the best solution at one step is not the best one for the entire training process.

Nevertheless, we add a new stage at the end of each iteration of the conventional SA, a new aggregation round, to assess the best solution (lines 18 --- 21). Evaluating the best solution allows adding randomness whether the best solution degrades, avoiding local minima. We only decrease the temperature parameter when a worse solution is accepted. The temperature is responsible for accepting worse solutions as a candidate to best solution.

The SA has an additional loop that is responsible for temperature reduction. For FedSA, we replace the inner loop with a new temperature reduction method. In our proposed variant, the temperature is only reduced when a worse solution is selected (lines 13 --- 17). Therefore, the probability $ P (\Delta E, T) $ of acceptance of a worse solution is computed at each FedSA iteration (line 8). The idea is enabling a trade-off between diversity and greed. Hence, the algorithm tends to be greedy only when it accepts generalist solutions.

\begin{figure}[h!]
\vspace{-3mm}
    \centerline{\includegraphics[width=.5\textwidth]{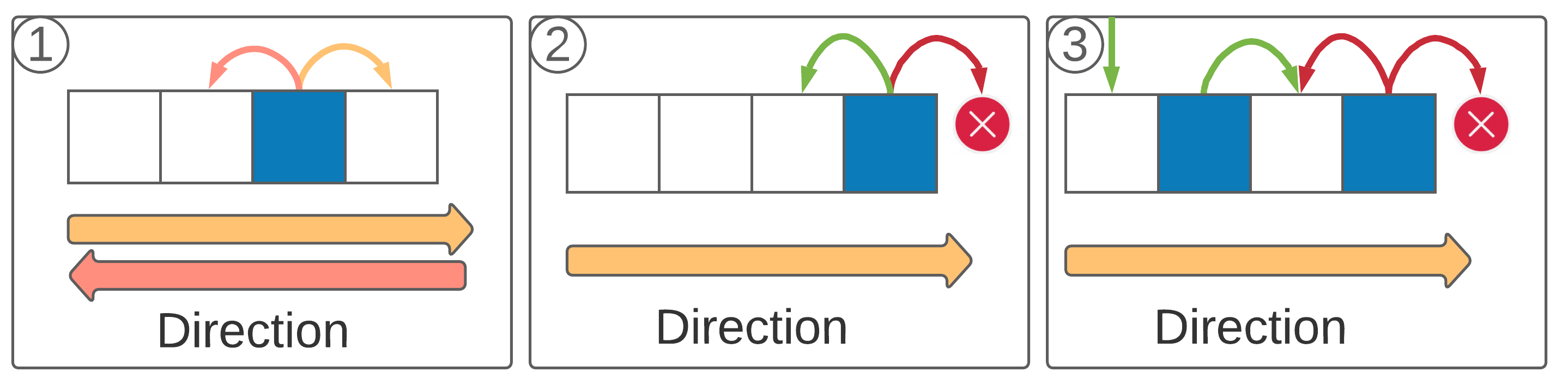}}
    \caption{Neighbor structure for integer values. The tick orange arrow represents the positive direction, and the salmon represents the negative direction. The thin green arrows are acceptable permutations, and the red ones are forbidden.}
    \label{fig:intneigh}
\end{figure}

The function \texttt{INIT} is responsible for generating a random candidate solution following a uniform distribution. The probability density function is $ \frac{1}{b - a} $, where $ a $ and $ b $ are lower and higher thresholds values respectively. The function \texttt{AGGREGATE} represents a global aggregation round and returns the loss of the global model. The function \texttt{GEN\_NEIGHBOR\_SOLUTION} generates a neighbor solution of the current best solution known. The generation of the best solution's neighbor depends on the variable type. The number of local updates is an integer value. 
Hence, our function selects a subsequent value of the current value as neighbor. 
Figure~\ref{fig:intneigh} (1), (2), and (3) are a graphical representation of the neighbor selection structure. Each square is an integer value that represents a number of local updates. The blue square represents the so-far best value. 
The scenario in Figure~\ref{fig:intneigh} (1) represents a normal situation, where the metaheuristic selects neighbor value to the current best value. The subsequent value may be in a positive direction, simply by adding $ +1 $ to the current value, or in a negative direction, reducing $ -1 $ to the current value. The second scenario is when the current best value is the first or last in the list, \textit{i.e.}, a hyperparameter's maximum or minimum value. In this case, the direction of the current iteration would cause the neighboring solution to extrapolate the border. Then, under this condition, the searching direction is flipped. Figure~\ref{fig:intneigh} (3) shows a scenario where flipping the direction leads to an already selected position, e.g., an already selected participant ID. In this case, a random participant is selected. This neighbor structure resembles a circular list.

The participant subset selection follows a similar structure. Each participant has an integer number as an ID, and selecting a neighbor means incrementing or decrementing the current ID as shown in Figure~\ref{fig:intneigh} (1) and (2). Nevertheless, as we are selecting a new set of participants, our proposed method could cause a participant to be selected more than once. In this case, the function performs one from two actions: the first is to flip the direction, as shown in Figure~\ref{fig:intneigh} (2), and the second is randomly choose another not yet selected participant. The function will first choose to flip the direction, but if the participant ID of the flipped direction has already been selected, it randomly selects another participant. 


The learning rate is a float value, requiring a different method to calculate a neighbor value. Hence, we use 

\begin{equation}
    \eta \gets \eta_{\beta} \pm \epsilon \times unif\left ( \eta_{a}, \eta_{b} \right ),
\end{equation}

\noindent where, $ \eta $ is the decimal hyperparameter, $ \eta _{\beta} $ is the best value so-far, $ \eta_{a} $ and $ \eta_ {b} $ are the smallest and largest values that the hyperparameter assumes. Then, a value is generated following a continuous uniform distribution between the limits values of the hyperparameter. The value is multiplied by a constant $ \epsilon $, where $ 0 < \epsilon < 1 $. The constant $ \epsilon $  is the step for the neighboring value: the larger the constant, the greater the step. The product of the uniform distribution and $ \epsilon $ is added or subtracted, depending on the iteration direction.

After running the neighborhood calculation, we set the candidate solution as a set composed of: i)~The selected participants subset  $ S_ {t} $; ii)~the learning rate $ \eta $; and iii)~the number of local updates $ \tau $. Then, we calculate the loss function for the new candidate solution (line 6) and we evaluate whether the new solution should replace the previous solution (lines 7-17).

At the end of every FedSA iteration, an aggregation round is performed (line 18). The new aggregation validates the best solution known. In the case of the known solution being no longer the best, a completely random solution is generated with the function \texttt{INIT} (line 20) to prevent the algorithm from being stuck in a single region. The best solution may no longer be the best, \textit{e.g.}, the best learning rate may change with some iteration, or a participant's dataset may no longer reduce loss. Thus, the FedSA constantly evaluates the best solution to avoid getting stuck with a solution no longer the best.

\section{FedSA Evaluation and Results}
\label{sec:results}

We compare the FedSA with FedAVG~\cite{BrendanMcMahan2017} as the baseline to evaluate our proposal. The primary purpose of the evaluation is to assert the convergence time reduction of FedSA upon the baseline proposal, FedAVG. We develop a Python-based simulator that generates the participants' processes and shares the dataset among all participants. The simulator gets a dataset and splits it into training and evaluation sets. It is essential to notice that the training and validation sets are balanced for normal and attack samples. The training set contains 70\% of the dataset, while the evaluation set contains 30\% of the samples. The simulator randomly splits the training set in shards to the $ n $ participants. Since the shards are composed of random samples selected from the training set, it is not guaranteed to contain attack samples. All the shards have the same size and represent the participants' local datasets. Each participant trains their local model using their local datasets without sharing data. Thus, each participant is unaware of the other participants' data.
At each aggregation round, the selected participants send the parameters of their model for aggregation, Equation\ref{eq:fedavg}, generating the global model $ w_G^t $. The aggregation server uses the validation set to calculate the global loss. The simulations were performed on a computer equipped with an Intel (R) Xeon Phi (TM) CPU 7250 @ 1.40GHz processor and 128GB of RAM.

We employ TensorFlow\footnote{ https://www.tensorflow.org/library} to build the machine learning models. The simulator implements FedAVG and the FedSA proposal. Our machine learning model is a Multilayer Perceptron (MLP), deploying two hidden layers. The first hidden layer contains 50 neurons, and the second, 100 neurons. We apply Rectified Linear Units (ReLU) as the activation function in hidden layers and Softmax in the output layer. We empirically choose the neural network configuration performing fine-tuning with the entire dataset in a centralized machine learning evaluation.


We evaluate the proposals against the  CICIDS2017 dataset~\cite{datasetids} to generate traffic on each local network. The CICIDS2017 dataset is from Canadian Institute for Cybersecurity (CIC), and contains both normal and attack flows. The CICIDS2017 is a dataset containing a variety of common network attacks. The CICIDS2017 testbed consists of a firewall and twelve hosts, and the attackers are located in a separate network.
The CICIDS2017 dataset contains five days of captured traffic and eight network attacks, including Brute Force FTP, Brute Force SSH, DoS, Heartbleed, Web Attack, Infiltration, Botnet, and DDoS. The CICIDS2017 dataset is unbalanced, containing around 80\% of normal flows and 20\% of attack flows. The dataset has 2,830,743 samples, which we have split into 70\% for training, and 30\% for testing.
Besides, we apply CICFlowMeter\footnote{Available at https://www.unb.ca/cic/research/applications.html} to transform network packets into bidirectional network flows. 
The CICFlowMeter tool generates 80 different features such as flow duration, number of packets, number of bytes, and average packet size. 
To avoid overfitting, we discard source and destination IP addresses, source and destination ports, and transport protocol. We used the same evaluations employed in previous works, with a plus of evaluating other metrics than only accuracy~\cite{preuveneers2018, Nguyen2019, Chen2020ids, rey2021}. We also evaluate the impact of selecting a small proportion of selected participants in each aggregation round.



\subsection{Simulation scenario}


We evaluate the accuracy, precision, sensitivity, specificity, f1 score, and loss function to measure the performance of FedSA and FedAVG. We use a scenario with 100 participant IDSes, and only 30\% of the participants are selected at each aggregation round. 
The proportion of selected participants affects the learning. Besides evaluating the metrics above,  we measure whether the proportion of selected participants affects the classification performance. 

We simulate two scenarios varying the proportion of selected participants to evaluate the impact on training. The main goal is to evaluate the impact of the chosen participants' ratio on the global model's performance and convergence. It is important to note that aggregating the parameter of all participants is not scalable since the growth in the number of participants may negatively impact the server processing capacity. The first scenario is a federation with 100 participants, in which we select 50 for the training ($ S^{t} $), 50\% of the whole set of participants. In the second scenario, we increase the number of participants to 150, and 40 participants are selected for training, changing the proportion of $ S^{t} $ from 50\% to $ \approx $ 27\%. Changing the number of selected participants, we aim to assess the impact of the participants' selection ratio on the proposals' convergence. The idea is to select a large and small size of selected participants.

FedAVG selects the same number of participants as FedSA, albeit at random. FedAVG employ a learning rate of $ \eta = 0.1 $ with a decay at each iteration of $ \gamma = \frac{0.1}{i} $, in which $ i $ is the number of the iteration. For the best of our knowledge, there is no formal study for the learning rate value; a fine-tuning step or empirical choice often occurs. On the other hand, previous proposals aim to improve the convergence of the neural network using adaptive learning rates~\cite{yu1995dynamic, yedida2021}. At each iteration, the learning rate updates to:

\begin{equation}
    \eta^t \gets \eta^{t-1} \times \frac{1}{1 + \gamma  i}. 
\end{equation}


For the FedAVG evaluation, the learning rate dynamically decrements during the communication rounds according to the decay. In FedSA, the learning rate adaptively changes during the training, which means that the learning rate value depends on the previous learning rate value. The only hyperparameter that remains static for FedAVG is the local update. Then, we evaluate scenarios with different local updates number.

Another crucial point is to evaluate whether the FedSA parameters impact the training. We evaluate the FedSA with different combinations of $ T $ and $ \alpha $ to evaluate the performance impacts of the FedSA hyperparameters. We also compare federated learning against centralized learning strategies to evaluate the overhead added by the federation procedure. The comparison with centralized machine learning is to evaluate the tradeoff between performance and privacy. The main focus is to highlight the performance losses due to privacy guarantees provided either by FedAVG or FedSA. 

\subsection{FedSA Evaluation}
\label{subsec:cicids}

\begin{figure}[h!]
\vspace{3mm}
\begin{center}
\mbox{
\subfigure[Comparison of the FedSA and FedAVG accuracy. ]{
\includegraphics[width=0.43\columnwidth]{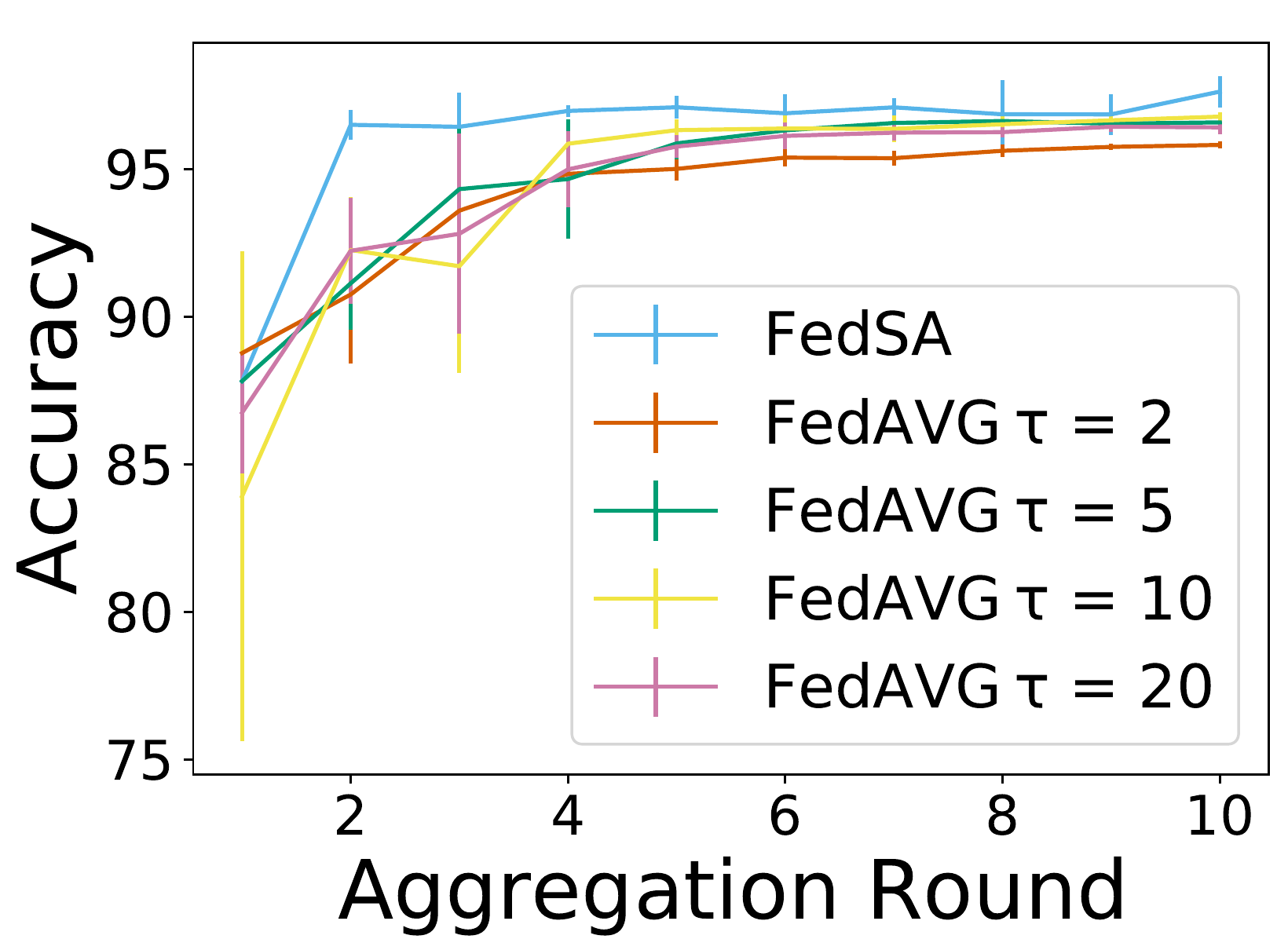}
\label{fig:accuracy1}
}}
\mbox{
\subfigure[Cross-entropy loss function evolution over the aggregation rounds.] {
\includegraphics[width=0.45\columnwidth]{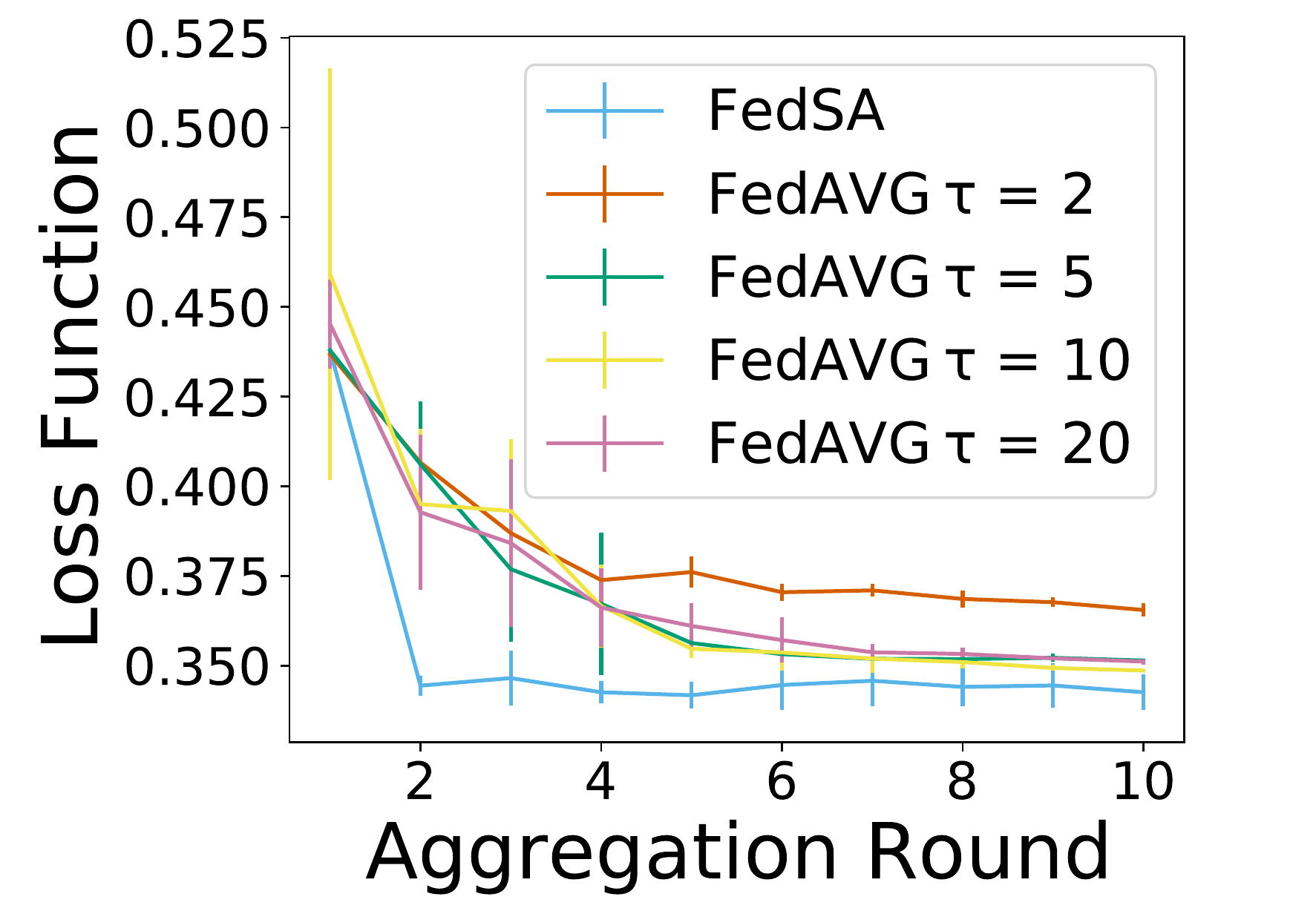}
\label{fig:loss1}
} }
\end{center}
\vspace{-3mm}
\caption{Accuracy and loss regarding the validation dataset. 
The acceleration in the convergence is related to the FedSA selection of the hyperparameters and participants.}
\vspace{-3mm}
\label{fig:result1}
\end{figure}

First,  we evaluate the impact of the local updates over the aggregation rounds of FedAVG compared against the FedSA proposal. We perform ten aggregation rounds because neither FedSA nor FedAVG improves performance after the tenth round. FedSA chose the number of local updates at each aggregation round in the interval of 1 to 20 local updates. The accuracy and loss of all evaluation was performed with the validation set. 
Figure~\ref{fig:result1} shows that the proposal reaches 95,5\% of accuracy in 2 rounds, while, to achieve the same accuracy, FedAVG takes eight aggregation rounds with ten local updates. 
The randomness of FedAVG while choosing the participants' subset for training may leads to the selection of participants that do not contribute to the model. Besides, the gradual reduction of the learning rate delays the convergence of the global model. The adaptive number of the local updates is essential to the global models' convergence since the need for local updates may diverge for each dataset and sometimes for each aggregation round. We also observe, Figure~\ref{fig:loss1}, that global loss significantly reduces in the first two iterations using FedSA, followed by minor changes in the subsequent rounds. Then, we assume that three aggregation rounds are a fair trade-off between classification performance and processing for a real-world environment.

\begin{figure}[t!]
\begin{center}
\mbox{
\subfigure[Varying participants' selection ratio for the model training.
]{
\includegraphics[width=0.44\columnwidth]{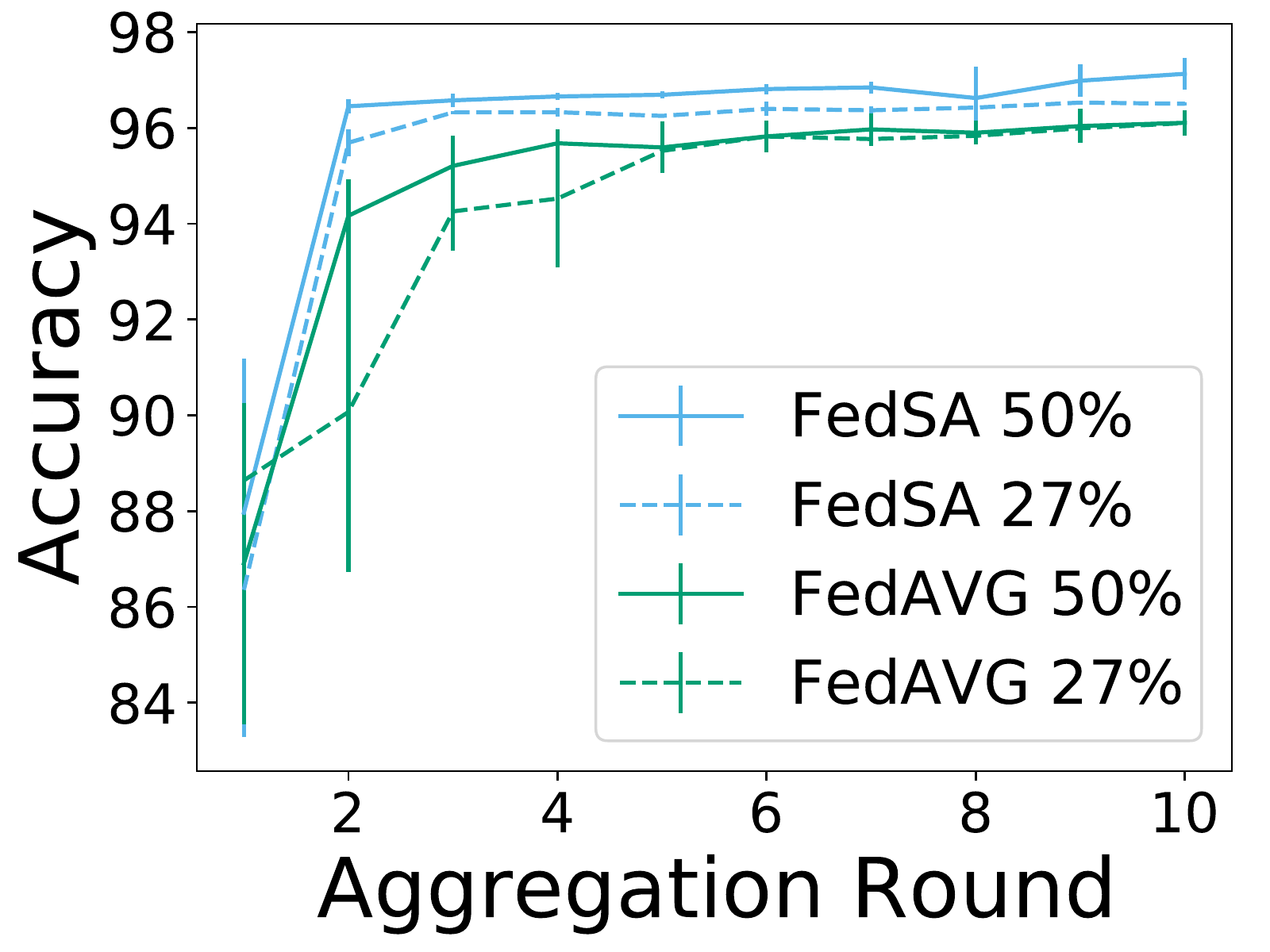}
\label{fig:acc_pro}
}}
\mbox{
\subfigure[Loss function evolution over the aggregation rounds.] {
\includegraphics[width=0.44
\columnwidth]{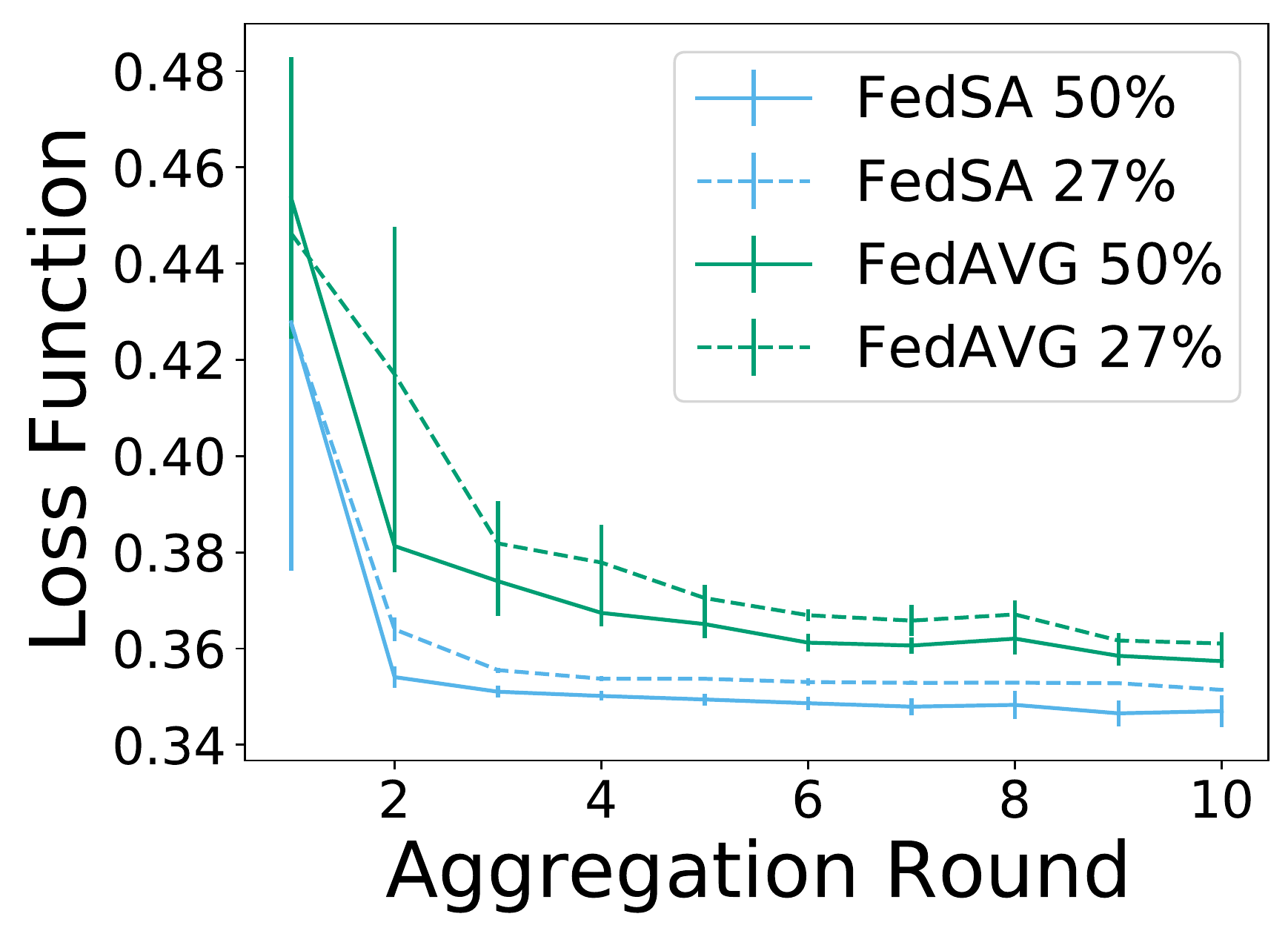}
\label{fig:loss_prop}
} }
\vspace{-3mm}
\end{center}
\caption{The global model performance for different participants' selection proportions. Comparison of two scenarios: 150 participants and 40 selected at each aggregation round (27\% selection scenario);  100  participants and 50 selected at each aggregation round (50\% selection scenario). }
\label{fig:result2}
\end{figure}

Figure~\ref{fig:result2} shows the proposed FedSA and the baseline FedAVG in two different scenarios. The first scenario is a federation with 100 participants and, at each aggregation round, the server selects 50\% for the training process. The second scenario is a federation with 150 participants, but just 27\% join the training process. The main focus of the test is to certify if selecting a different proportion of participants affects the convergence of the model. For the sake of fairness, FedAVG applies the number of local updates that best performed in previous evaluations. Figure~\ref{fig:result1} shows that ten local updates work better than two, five, and twenty local updates for FedAVG using the CICIDS2017 dataset.
Selecting fewer participants does not affect the convergence, as shown in Figure~\ref{fig:result2}. It is important to note that the FedSA achieves $ \approx 97\% $ accuracy even selecting a small proportion of participants in just three aggregation rounds. In comparison, FedAVG needs eight aggregation rounds. In addition, Figure~\ref{fig:acc_pro} shows that the scenario with fewer participants (27\%) obtains better accuracy than the scenario in which 50\% of the participants are selected to train the global model. We hypotheses that selecting fewer participants increases the classification performance, but it is still crucial to consider other metrics, such as precision, sensitivity, specificity, and f1 score.

\begin{figure}[h!]
\begin{center}
\mbox{
\subfigure[The global model precision, sensitivity, specificity, and f1 score in a federation with 100 participants and 50 participants selected to $ S^{t} $ subset.
]{
\includegraphics[width=0.44\columnwidth]{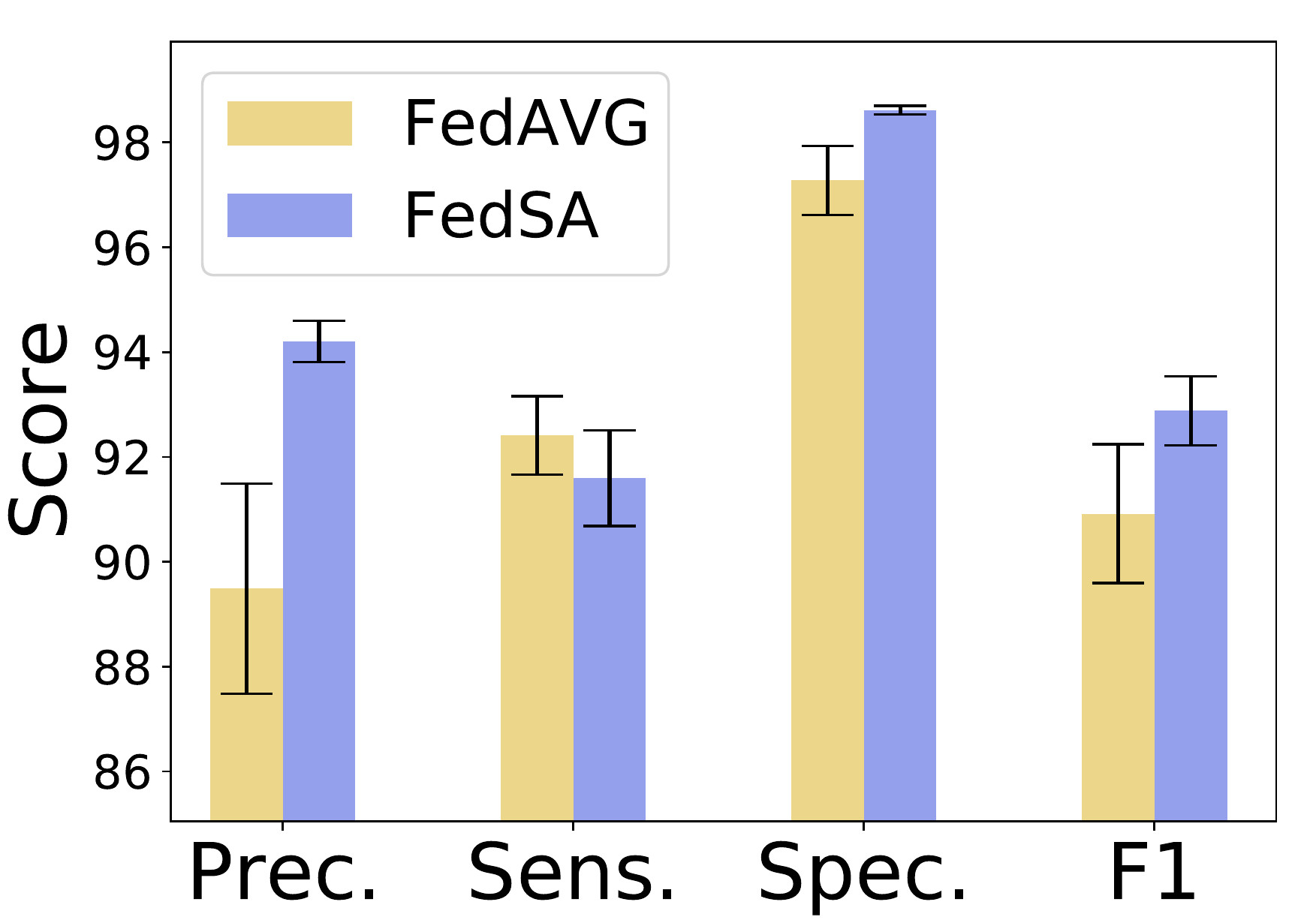}
\label{fig:bar1}
}}
\mbox{
\subfigure[The global model precision, sensitivity, specificity, and f1 score in a federation with 150 participants and 40 participants selected to $ S^{t} $ subset.] {
\includegraphics[width=0.43
\columnwidth]{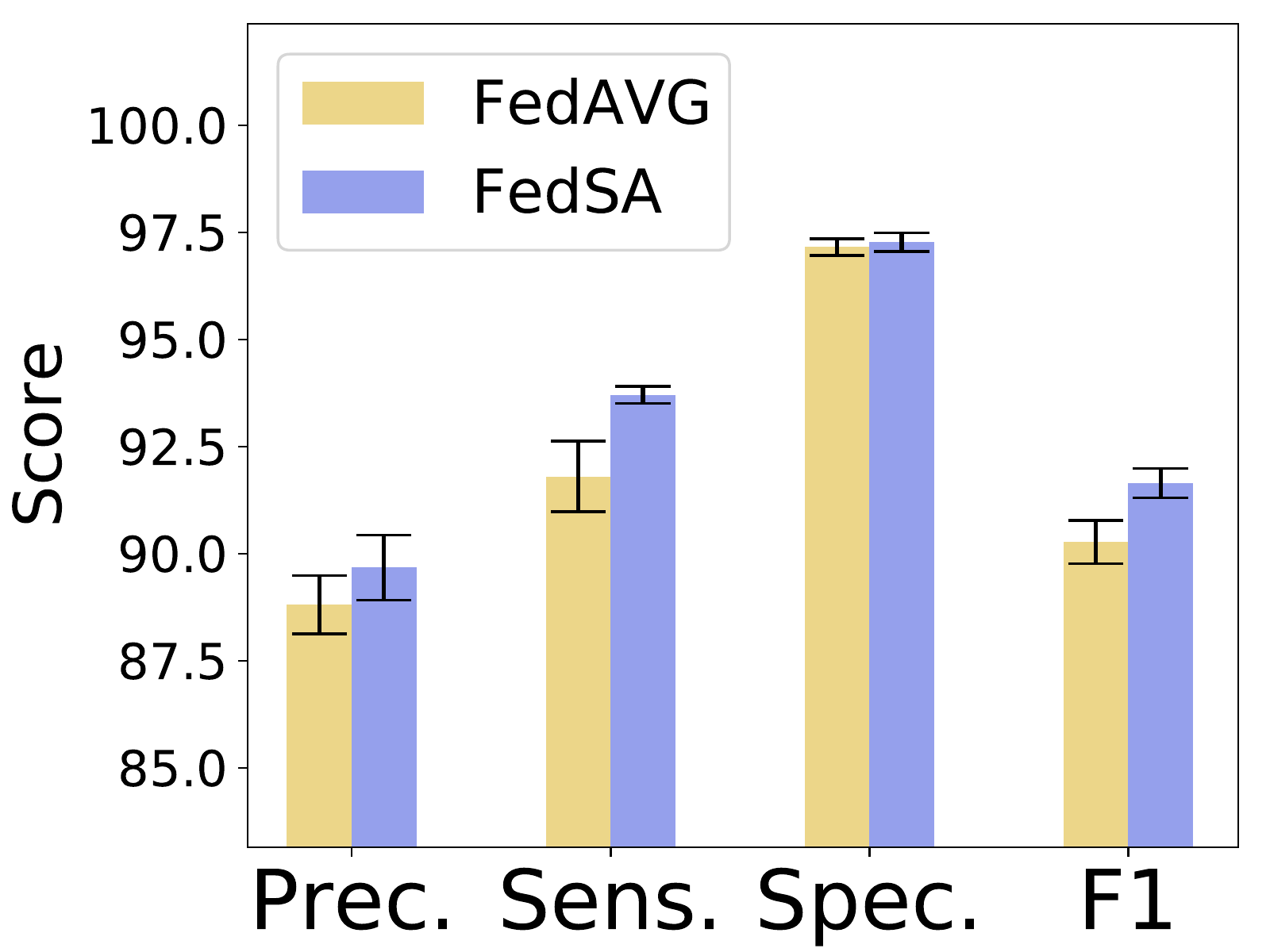}
\label{fig:bar2}
} }
\end{center}
\caption{The global model metrics for FedSA and FedAvg in the scenario of 50\% and $ \approx $ 27\% of participants' subset selection. (Prec.: Precision, Sens.: Sensitivity, Spec.: Specificity)}
\label{fig:bars}
\end{figure}

The validation dataset, similar to the participant's local dataset, is also unbalanced. There are other metrics such as precision, specificity, sensitivity, and f1 score for unbalanced datasets to evaluate if the model has good predictions for all classes. Precision evaluates how many attack samples the model predicted correctly.  Sensitivity, or Recall, measures the proportion of actual attacks cases predicted as an attack by the model. On the other hand, specificity measures the proportion of actual normal traffic predicted as normal traffic by the model. The f1 score combines precision and sensitivity by taking their harmonic mean. Figure~\ref{fig:bars} shows the precision, sensitivity, specificity, and f1 score for FedSA and FedAVG in two scenarios. The first scenario, shown in Figure~\ref{fig:bar1}, has 100 participants, and 50 participants are selected at each aggregation round to train the global model. FedSA presents better precision and specificity, but worst sensitivity, \textit{i.e.}, the FedAVG detected more attack samples in the validation dataset than FedSA. However, FedSA had better precision, \textit{i.e.}, our proposal presents fewer false-positive samples than FedAVG. False-positive is a significant problem for the IDS scenario since classifying a normal flow as attack flow will cause undesirable false alarms.

The evaluations show that selecting fewer participants for the training does not compromise the convergence, but it reduces the model's efficacy. The global model performs reasonable specificity in both scenarios, showing that the proposed algorithm correctly detects normal flows.

Another essential evaluation is a comparison of federated learning and centralized machine learning. For the evaluation, we measure the accuracy and the loss function of centralized machine learning, FedSA, and FedAvg. Centralized machine learning performs better because the model training accesses the entire dataset. A centralized machine learning approach implies that all the participants' data is centralized in a single dataset for training a single machine learning model. Besides, while training a collaborative model, the data sharing may add some noise to the training. 
Each aggregation round is equivalent to ten iterations of centralized machine learning. We used ten iterations per aggregation round because we use ten local updates for FedAvg.

\begin{figure}[h!]
\begin{center}
\mbox{
\subfigure[Accuracy comparison among centralized machine learning, FedSA and FedAVG.]{
\includegraphics[width=0.44\columnwidth]{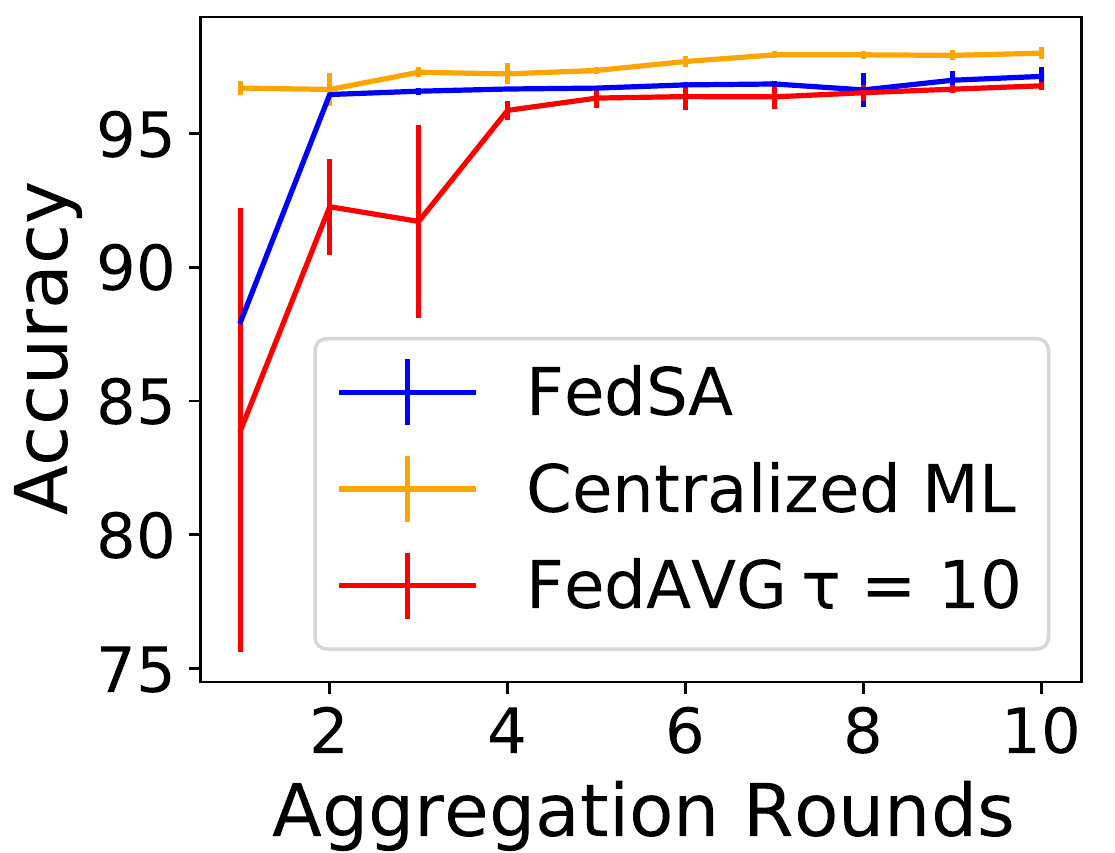}
\label{fig:centralized:acc}
}}
\mbox{
\subfigure[The loss function for the centralized and the federated learning global models.] {
\includegraphics[width=0.44
\columnwidth]{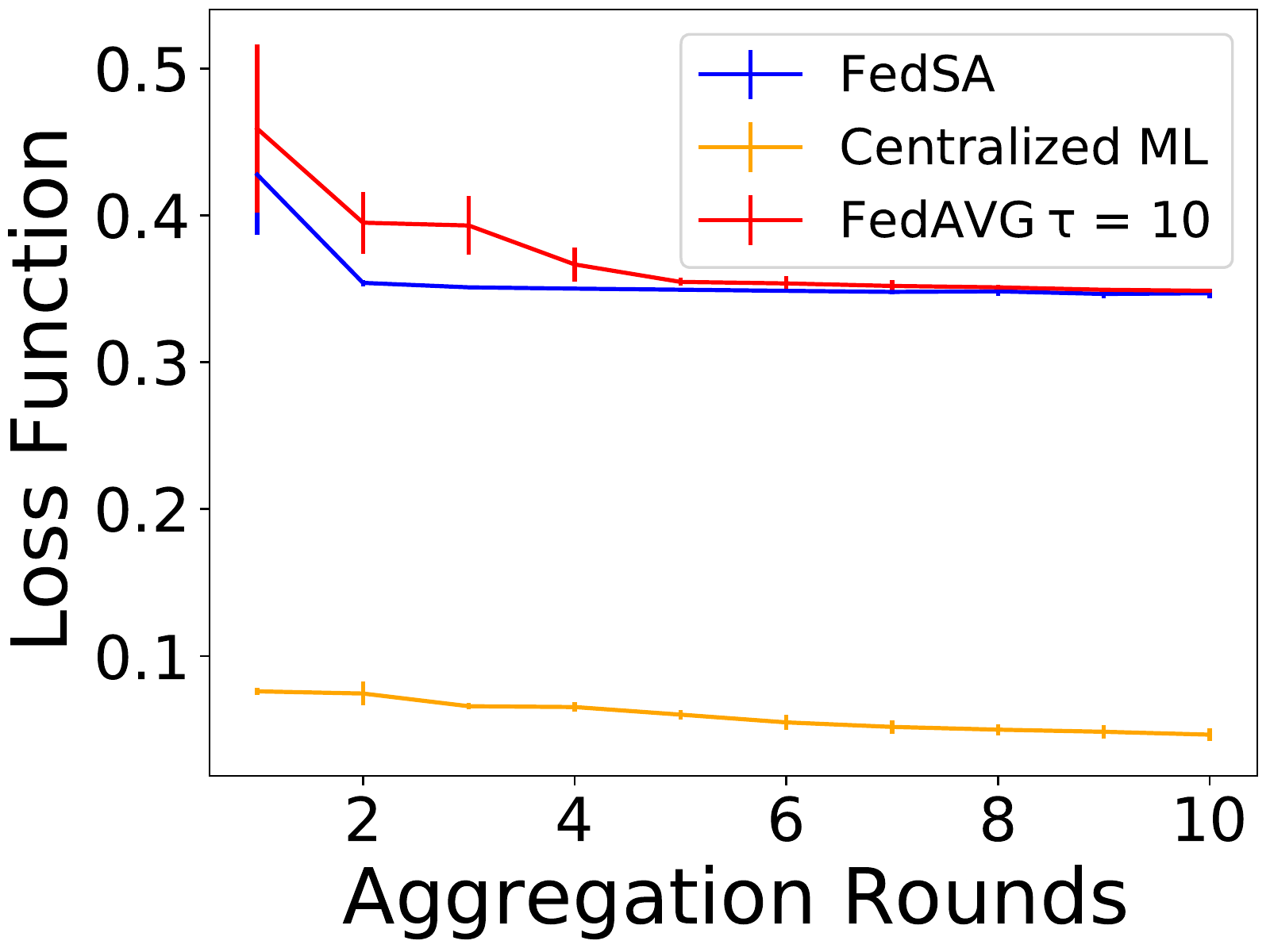}
\label{fig:centralized:loss}
} }
\vspace{-3mm}
\end{center}
\caption{For centralized machine learning (non-federated learning), one aggregation round is equal ten epochs. As expected, centralized learning is slightly better than the federated learning approach. }
\label{fig:centralized}
\end{figure}

Figure~\ref{fig:centralized} shows that the federated learning global models achieve an accuracy value close to one from the centralized machine learning model. Although centralized machine learning performs better than federated learning, the centralized approach relies on collecting all samples from the participants' dataset, harming participant privacy. It is important to highlight that the accuracy does not change much, even with a significant loss difference between the centralized approach and the federated ones. This phenomenon happens because the dataset reached a saturation point, where even reducing plenty of the loss does not change the accuracy.

The FedSA metaheuristic has two hyperparameters, cooling ($ \alpha $) and temperature ($ T $). The temperature determines the acceptance probability of a solution worse than the best. High temperatures lead to the acceptance of random solutions, and low temperatures lead to greedy behavior, refining an already good and accepted solution. The cooling hyperparameter is responsible for decreasing the temperature periodically. Hence, we used a high temperature $ 0.8 $ and a low cooling $ 0.05 $ in our evaluations. The following evaluation aims to assess the impact of the FedSA hyperparameters on the global model accuracy. This evaluation intends to show that the FedSA hyperparameters do not drastically change the global models' accuracy. We perform the FedSA hyperparameters evaluations using ten aggregation rounds.

\begin{table}[h!]
\caption{Evaluating FedSA with different cooling and initial temperature. 
}
\centering
\begin{tabular}{l|l|l|l|}
\cline{2-4}
                                     & T = 0,1                                                                    & T = 0,4                                                                    & T = 1                                                                       \\ \hline
\multicolumn{1}{|l|}{cooling = 0,05} & \begin{tabular}[c]{@{}l@{}}$\mu = 96.84 $\\ $ \sigma = 0.08 $\end{tabular} & \begin{tabular}[c]{@{}l@{}}$\mu = 96.74 $\\ $ \sigma = 0.02 $\end{tabular} & \begin{tabular}[c]{@{}l@{}}$\mu =  96.66 $\\ $ \sigma = 0.01 $\end{tabular} \\ \hline
\multicolumn{1}{|l|}{cooling = 0,4}  & \begin{tabular}[c]{@{}l@{}}$\mu = 96.71 $\\ $ \sigma = 0.04 $\end{tabular} & \begin{tabular}[c]{@{}l@{}}$\mu = 96.74 $\\ $ \sigma = 0.09 $\end{tabular} & \begin{tabular}[c]{@{}l@{}}$\mu = 96.67 $\\ $ \sigma = 0.07 $\end{tabular}  \\ \hline
\multicolumn{1}{|l|}{cooling = 0,9}  & \begin{tabular}[c]{@{}l@{}}$\mu = 96.80 $\\ $ \sigma = 0.05 $\end{tabular} & \begin{tabular}[c]{@{}l@{}}$\mu = 96.61 $\\ $ \sigma = 0.10 $\end{tabular} & \begin{tabular}[c]{@{}l@{}}$\mu = 96.56 $\\ $ \sigma = 0.06 $\end{tabular}  \\ \hline
\end{tabular}
\\
$\mu$: mean; $\sigma$: standard deviation
\label{tab:fedsahp}
\end{table}

Table~\ref{tab:fedsahp} reveals combinations of FedSA hyperparameters. We perform tests for each combination several times for five aggregation rounds. Then, we calculated the mean ($ \mu $) and the standard deviation ($ \sigma $) of each combination. We deploy only five aggregation rounds because, in the previous evaluation, the FedSA global model, at the fifth round, achieved a similar result to the last aggregation round. Thus, we assume the global model convergence at the fifth aggregation round. 


Based on the results of Table~\ref{tab:fedsahp}, we show that the selection of FedSA input hyperparameters does not interfere the final results. It is essential to note that temperature values between 0.4 and 1 have a high probability of accepting a worse solution in our scenario. This phenomenon is because the difference between the best and current losses is generally tiny.  Then, we can affirm that $ T = 1 $ is a high temperature
Table~\ref{tab:fedsahp} also shows that even drastically changing the FedSA's hyperparameters, the model achieved $ \approx 97\%$ accuracy in 5 aggregation rounds. Therefore, FedSA hyperparameters require no fine-tuning process. Nonetheless, traditional Federated Learning hyperparameters, such as learning rate and local update number, drastically change the global model accuracy, requiring fine-tuning process. Then, we can affirm that using a uniform distribution function to select FedSA's hyperparameters would not harm the training.

\section{Conclusion}
\label{sec:conclusion}

Federated learning for Intrusion Detection Systems (IDS) shows promising results in detecting new attack patterns. Federated learning presents several optimization challenges since the aggregation server does not access training data. The traditional random participant selection approach may lead to selecting participants that do not contribute to the convergence. Also, the essential hyperparameters for the federated learning convergence, the number of local updates, and the learning rate have fixed values in the FedAvg algorithm. 

In this paper, we proposed the Federated Simulated Annealing (FedSA) to accelerate the convergence of the global model without the need for a time-consuming process of fine-tuning searching for the best set of hyperparameters. The search for the best hyperparameter is performed by the FedSA adaptively during the training. The results showed that FedSA optimizes the training achieving $ \approx 97\% $ accuracy using the CICIDS2017 dataset, while the baseline FedAVG took almost twice aggregation rounds to reach the same result. The results also showed that even after selecting a small proportion of the participants for training, the FedSA proposal achieved 96\% accuracy in just three aggregation rounds while the baseline needed eight. In addition, FedSA selecting 27\% of participants achieved 96\% accuracy in three aggregation rounds while FedAvg selecting 50\% of participants took eight aggregation rounds. The proposed solution achieves maximum accuracy with five aggregation rounds, 50\% faster than the traditional.

We will improve the neighbor structure for the participants' selection as future work. Currently, we use participants' user IDs for the neighborhood structure. Nevertheless, using participants with similar model parameters as neighbors is the goal. Another crucial point for future works is the benchmarking of FedSA with other proposals than FedAvg.

\section*{Acknowledgements}
This research is supported by CNPq, CAPES, RNP, and FAPERJ.


\balance

\bibliographystyle{IEEEtran}  
\bibliography{main}

\end{document}